\documentstyle[prl,aps,epsf]{revtex}
\begin{document}
\draft

\twocolumn[\hsize\textwidth\columnwidth\hsize\csname
@twocolumnfalse\endcsname

\title{Dynamic Phases of Vortices in Superconductors with Periodic Pinning} 

\author{C.~Reichhardt, C.~J.~Olson, and Franco Nori}
\address{Department of Physics, The University of Michigan,
Ann Arbor, Michigan 48109-1120}

\date{\today}
\maketitle
\begin{abstract}

We present results from  extensive simulations of driven vortex 
lattices interacting with periodic arrays of pinning sites.  
Changing an applied driving force produces a rich variety of 
novel dynamical plastic flow phases which are very distinct from 
those observed in systems with random pinning arrays. 
Signatures of the transition between these different dynamical 
phases include sudden jumps in the current-voltage curves as 
well as marked changes in the vortex trajectories and vortex 
lattice order.  Several dynamical phase diagrams are obtained 
as a function of commensurability, pinning strength, and spatial 
order of the pinning sites. 
\end{abstract}

\vspace{-0.25in} 
\vskip1pc]
\vskip2pc 
\narrowtext
The dynamics of driven vortex lattices interacting with 
quenched disorder have recently attracted considerable attention.
Both theoretical \cite{Brass,Koshelev,Giamarchi,Kyu,Middleton} 
and experimental \cite{Yaron} studies have suggested the exciting 
possibility that these systems exhibit novel dynamic phase 
transitions between different flow regimes as a function of 
driving force.
In samples with random pinning, evidence for a transition
from a plastic flow phase to a strongly driven ordered homogeneous 
phase has produced current debate over whether the strongly driven 
phase forms a moving crystal\cite{Koshelev} or moving ordered 
glass \cite{Giamarchi,Kyu}.  The dynamic phases may also be relevant 
to many other systems, such as Josephson junction arrays, 
charge-density waves, and electron crystals. 
Although much work has been done on dynamic phases in systems  
with {\it random} disorder, the case of {\it periodic} pinning has 
not been addressed. We show that driven vortex lattices interacting 
with periodic pinning exhibit a number of novel {\it plastic flow 
phases} which are not observed in random pinning arrays. 
Further, the onset of these different phases produces microscopic 
features in the vortex structure and flow patterns and gives rise 
to very pronounced features in macroscopically measurable 
current-voltage curves. 

General interest in periodic arrays of pinning sites has increased 
now that it is possible to construct samples with well defined 
periodic pinning structures in which the microscopic pinning 
parameters, such as size, depth, periodicity, and density, can be 
carefully controlled \cite{Rosseel,Harada,Reichhardt}. 
Interesting commensurability effects  are observed both in 
magnetization measurements \cite{Rosseel} and with direct 
imaging \cite{Harada}.   
Periodic pinning arrays are also of technological importance since 
they can produce higher critical currents than an equal number of
randomly placed pins \cite{Reichhardt}.

When the pinning radius is much smaller than the lattice spacing, 
the ``disorder'' in the system can be {\it fine tuned\/} by 
changing the commensurability. At $ B/B_{\phi} = 1$, where 
$B_{\phi}$ is the field at which the number of vortices $N_{v}$ 
 equals the number of pinning sites $ N_{p}$, the vortex lattice 
locks into a periodic pinning array \cite{Rosseel,Harada,Reichhardt} 
and the pinning force is maximized. For $B/B_{\phi} > 1$, the vortex 
lattice contains {\it two species of vortices}: the pinned vortices 
that are commensurate with the pinning  array, and the generally more 
weakly pinned {\it interstitial} vortices that are caged by vortices at
the pinning sites. When $ B/B_{\phi} < 1$, the vortex lattice contains 
a well defined number of {\it vacancies} which have their own dynamical 
behavior.  

In order to study the two-dimensional (2D) dynamics of rigid flux 
lines driven over periodic arrays of pinning sites, we have 
performed a large number of $T = 0$, current-driven molecular 
dynamics (MD) simulations.  Unlike previous current-driven 
simulations \cite{Brass,Koshelev,Kyu,Middleton}, we examine the 
effects of periodic pinning arrays rather than random arrays, 
and cover a much larger range of the microscopic pinning and 
system parameters, allowing us to construct a series of detailed 
dynamical phase diagrams.  

We numerically integrate the overdamped equations of motion\cite{rmp}:
$ {\bf f}_{i} = {\bf f}_{i}^{vv} + {\bf f}_{i}^{vp} + 
{\bf f}_{d} = \eta{\bf v}_{i}$. 
Here $ {\bf f}_{i}$ is the total force acting on vortex $ i $
and we take $\eta = 1$. The force from the other vortices is  
$ {\bf f}_{i}^{vv} = \ \sum_{j=1}^{N_{v}} \ f_{0}\, 
K_{1}(|{\bf r}_{i} - {\bf r}_{j}| / \lambda) \ 
{\bf {\hat r}}_{ij} $, 
where  $K_{1}(r/\lambda)$ is a modified Bessel function, 
$\lambda$ is the penetration depth, and
$ {\bf {\hat  r}}_{ij} = $ 
$ ({\bf r}_{i} - {\bf r}_{j})/|{\bf r}_{i} - {\bf r}_{j}|$. 
A cut off is placed on $K_{1}(r/\lambda)$ after it reaches an 
extremely small value at $r = 6\lambda$.  
The pinning force is 
$ {\bf f}_{i}^{vp} = $
$ \ \sum_{k=1}^{N_{p}}(f_{p}/r_{p})|{\bf r}_{i} - $ $ {\bf r}_{k}^{(p)}|
\ \Theta ( (r_{p} - $ $ |{\bf r}_{i} -{\bf r}_{k}^{(p)}|)/\lambda)
\,{\bf {\hat r}}_{ik}^{(p)}.$ Here, $ \Theta$ is the step function,
$ {\bf r}_{k}^{(p)}$ is the location of pinning site $k$, 
$f_{p}$ is the maximum pinning force,  and 
$ {\bf {\hat r}}_{ik}^{(p)} = $
$ ({\bf r}_{i} - $ $ {\bf r}_{k}^{(p)})/|{\bf r}_{i} - $ 
$ {\bf r}_{k}^{(p)}| $. 
The Lorentz force is modeled as a uniform force  $ {\bf f}_{d}$. 
All lengths, fields and forces are given in units of 
$ \lambda$, $\Phi_{0}/\lambda^{2}$, and 
$ f_{0} = \Phi_{0}^{2}/8\pi^{2}\lambda^{3}$, respectively. 
 
We focus on experimentally accessible parameters that are 
close to those used in recent experiments \cite{Rosseel}. 
Fixing the sample size at $36\lambda\times 36\lambda$, 
with periodic boundary conditions, and the pinning radius 
at $ r_{p} = 0.3\lambda$, we examine an $ 18\times 18$ 
square pinning array with $ N_{p} = 324$, giving a pinning 
density of $ n_{p} =  0.25/\lambda^{2}$. 
We slowly increase the driving force $f_{d}$ along the 
horizontal symmetry axis ($x$-axis) of the pinning lattice 
and compute the average velocity in the $x$-direction, 
$ V_{x} = (1/N_{v})\sum_{i=1}^{N_{v}}{\bf v}_{i}\cdot{\bf {\hat x}}$.
This quantity is proportional to a macroscopically measured 
voltage-current $V(I)$ curve.  
In order to separate the different effects that each of the 
pinning and system parameters have on the vortex dynamics, we 
fix all the parameters and vary only one at a time. 

\begin{figure}
\centerline{
\epsfxsize=2.8in
\epsfbox{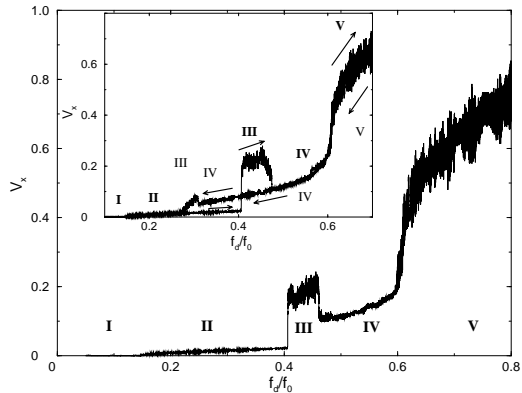}}
\caption{ 
Average vortex velocity $V_{x}$ versus driving force $f_{d}$,
for $B/B_{\phi} = 1.062$, $f_{p} = 0.625f_{0}$, 
$r_{p} = 0.3\lambda$, and $B_{\phi} = 0.25\Phi_{0}/\lambda^{2}$,
with the pinning sites located in a square array; $f_{d}$ is 
increased from $ 0 $ to $ 0.8f_{0}$. Several remarkable jumps 
in the curve can be clearly seen which correspond to transitions 
in the dynamical behavior of the driven lattice.
To better identify the  phases we have numbered them I--IV.
The inset shows the hysteresis curve as $f_{d}$ is increased to 
$ 0.7f_{0}$ and then decreased (phases now shown with unbold 
Roman numerals) to zero. Some phase boundaries, II--III and 
III--IV, show hysteresis while others do not. 
}
\label{fig1}
\end{figure}
In Fig.~1 we present a typical $V(I)$ curve for $ B > B_{\phi}$ 
as the driving force is linearly increased from $0$ to
$ 0.8f_{0}$ with $f_{p} = 0.625f_{0}$, 
$B_{\phi} = 0.25\Phi_{0}/\lambda^{2}$, and $ B/B_{\phi} = 1.062$. 
The $V(I)$ curve exhibits several remarkable features which 
clearly appear as {\it discontinuous} jumps and drops in 
$ V_{x}$. We label each of the features as regions I through V,   
and approximate the fraction of flux lines which are mobile at a 
specific driving force with: $ \sigma = V_{x}/f_{d}$.
As $f_{d}$ is increased in region I, $V_{x}$ is zero,
indicating that the vortex lattice is pinned. At $f_{d} = 0.146f_{0}$,
the onset of region II is marked by a finite $V_{x}$, caused by 
the depinning of interstitial vortices. Here, $\sigma = 0.06$, 
confirming that only the interstitial vortices are mobile 
since the percentage of vortices above $ B_{\phi}$ is also 
$ (B - B_{\phi})/B_{\phi} = 0.06$.  Region III begins at  
$ f_{d} = 0.406f_{0}$, where a very sharp jump up in $ V_{x}$ 
is seen, along with an {\it increase} in the number of mobile 
vortices to $ \sigma = 0.44$.
The velocity fluctuations $\delta V_{x}$ are also much larger. 
At $f_{d} = 0.462f_{0}$, region IV appears with a 
{\it sudden drop\/} in $V_{x}$, with $\sigma = 0.23$ and a 
reduction in $ \delta V_{x}$.  
Finally, at $ f_{d} = 0.612f_{0}$, just under the pinning 
force of each pinning site ($ f_{p} = 0.625f_{0}$), the 
entire lattice becomes mobile with $ \sigma = 1$, and the 
system enters region V. 

To further characterize the dynamic phases, we have performed 
a number of hysteresis runs where the driving force is slowly 
increased and then decreased. In the inset of Fig.~1 we show a 
typical curve for the system with the same parameters as in 
Fig.~1. There is little hysteresis for the transition IV--V.  
However, there is a very strong hysteresis at the transitions 
II--III and III--IV that persists in larger system sizes and 
also for low values of the spatial disorder. The hysteresis 
and the sharp jumps suggest that the II--III and III--IV 
phase boundaries might be first order.  Details will be 
presented elsewhere \cite{CR}. 

We use a series of snapshots of the vortex position and 
vortex flow paths for regions II through V to show 
explicitly that the features in the $V(I)$ curve correspond 
to {\it different plastic flow phases}. 
Figure 2(a) shows the vortex trajectories in region II of the 
current-voltage curve from Fig.~1.  It is clear that in region 
II only the interstitial vortices are mobile while the 
commensurate vortices remain pinned.  The motion is confined 
to {\it 1D channels} between the rows of pinning sites due to 
the square symmetry imposed by the pinned vortices. 
Such flow behavior for vortices in square pinning arrays has 
recently been experimentally observed \cite{Harada}.  

Figure 2(b) illustrates that the vortex trajectories differ
greatly from region III to II.  The vortex lattice is now 
disordered, and the flow is no longer 1D but consists of 
channels that wind in {\it both} the $x$ and $y$ directions. 
Pin-to-pin motion also appears. Unlike the motion in region 
II, where only the interstitial vortices move and vortices 
at the pinning sites remain pinned, {\it all} the vortices 
in region III take part in the motion, with any one vortex 
moving for a time and then being temporarily trapped \cite{web}. 

Another significant change in the vortex motion appears when 
the system enters region IV, as seen in Fig.~2(c).  
The vortex trajectories become more ordered and return to an 
exclusively 1D flow, with the mobile vortices moving 
{\it along} the pinning rows rather than {\it between} 
the rows as in region II. Only certain rows are mobile, 
and in these rows  the additional vortices above $ B_{\phi}$  
leave their positions between the pinning rows to create moving 
{\it incommensurate 1D structures} along the pinning row.
An entire row does not move continuously, but instead a pulse 
appears in which only four vortices, near the incommensurate 
segment of the vortex row, are mobile.
As this pulse or discommensuration moves across the sample, 
each vortex in the row is displaced by a single pinning 
lattice constant $a$.  This disturbance is thus crossing 
the sample much more rapidly than the vortices themselves.
 
The vortex trajectories  for region V, in which the entire 
lattice is moving, are shown in Fig.~2(d).
Some portions of the vortex lattice have a distorted triangular 
order, although the incommensurabilities from region IV are 
still present. 
The flow remains strictly 1D and along the pinning rows as in  
region IV, except that now {\it all} the rows are mobile. 
Rows with an incommensurate number of vortices  move faster 
than the commensurate rows. 
As $ f_{d}$ is increased further, the density of 
incommensurabilities and the vortex lattice structure do not 
change, so the system is {\it always undergoing plastic flow} 
and a moving crystal is not formed.   

The onset of these different phases is described with force 
balance arguments that take into account the coupling of 
the two different species of vortices, interstitial and 
(pinned) commensurate. 
In region II, while the commensurate vortices remain pinned,
the interstitial vortices begin flowing at a well defined 
driving force and exert a force $ {\bf f}_{c-ic}$ on the pinned 
commensurate vortices. 
The total force  on a commensurate vortex thus consists of
the driving force $ {\bf f}_{d}$,  pinning force $ {\bf f}_{p}$, 
and  forces ${\bf f}_{c-c}$  from commensurate and 
$ {\bf f}_{c-ic}$  interstitial vortices. The commensurate vortex 
will remain pinned as long the following inequality holds: 
\begin{equation} 
|{\bf f}_{p}| >| {\bf f}_d  + {\bf f}_{c-ic} + {\bf f}_{c-c}| \, . 
\end{equation}
Since here we are using a square pinning lattice, from symmetry 
we have ${\bf f}_{c-c} = 0$. If there are no incommensurate 
vortices, ${\bf f}_{c-ic} = 0$, and commensurate vortices depin at 
$|{\bf f}_{d}| = |{\bf f}_{p}|$. When interstitial vortices are 
present, the term ${\bf f}_{c-ic}$ causes commensurate 
vortices to depin before $ | {\bf f}_{d}| = |{\bf f}_{p}|$. 
These vortices  depin more vortices so that the number of mobile 
vortices increases and region III appears.  As long as 
$ | {\bf f}_{d}| < |{\bf f}_{p}|$, not all the vortices will be 
mobile so that $\sigma < 1$.  
For the parameters used in Fig.~1, the density  of interstitial 
vortices is sufficiently low that they do not interact 
significantly with each other. In this case, we can solve 
Eq.~(1) for the transition from region II to region III to give

\begin{equation}
f_{p} = 
\left[f_{d}^{2} + f_{0}^{2}K_{1}^{2}
\left(\frac{a + r_{p}}{2\lambda}\right)\right]^{1/2} \, . 
\end{equation} 
With the parameters for the onset of region III, Eq.~(2) gives 
$f_{p} = 0.621f_{0}$, which is in very good agreement with 
the value of $f_{p} = 0.625f_{0}$ used in the simulation.  

The appearance of 1D motion exactly along the rows of pins 
in region IV might seem counterintuitive since for 
$ B/B_{\phi} > 1$ and $ f_{d} = 0 $, when the vortices are 
not moving, an incommensurate vortex located along a pinning 
row is unstable to perturbations in the $y$-direction and will 
fall into the interstitial area between rows. 
For moving vortices, the situation is quite different since 
the vortices spend part of their time {\it in} the pinning sites.
The pinning sites create a stabilizing force against perturbations
in the transverse direction, confining the motion along the 
pinning rows.  When the density of interstitial vortices is 
low, the onset of region IV occurs when the driving force is 
strong enough that interstitial vortices can depin commensurate 
vortices from a distance $a/2$  in the longitudinal direction. 
This distance, and especially the repulsion from the remaining 
pinned vortices, allow the interstitial vortex to move towards 
the just-vacated pin site.  For sufficiently strong $f_d$, 
it will remain moving along the pinned rows \cite{CR}.
The transition from region III to region IV should occur at 
\begin{equation}
f_{p} = \left[\left(f_{d} +
\frac{f_{0}}{\sqrt 2}K_{1}\left(r_{1}\right)\right)^{2} 
+ \frac{f_{0}^{2}}{2}K_{1}^{2}\left(r_{1}\right)\right]^{1/2} \, , 
\end{equation}
where $ r_{1} = (a/\sqrt 2 + r_{p})/\lambda$. 
With the parameters used in Fig.~1 at the onset of the 1D incommensurate 
flow, Eq.~(3) gives $f_{p} = 0.624f_{0}$, which is in very good agreement 
with the numerical value for $f_{p}$ shown in Fig.~1. 
   
To better characterize the flow behavior, we  systematically vary 
$f_{p}$ with the rest of the parameters fixed.  
The resulting phase diagram in Fig.~3(a) outlines the onset of the 
different dynamical phases. As $ f_{p}$ is increased, region I 
saturates at a value of $ f_{d} \approx 0.146f_{0}$. This occurs 
because, although the pinning force $ f_{p}$ of the pinning sites 
is being increased, the vortex-vortex interactions which determine 
the interstitial pinning force are not changed.  
Region II only occurs when $ f_{p} > 0.37f_{0}$,  since for 
$ f_{p} < 0.37f_{0}$,  Eq.~(2) cannot be satisfied even for very 
low $ f_{d}$, and as soon as the interstitial vortices move they 
start to depin commensurate vortices.
The same argument applies for the onset of region IV, which 
extends to even lower $f_{p}$ values. The II--III phase 
boundary follows Eq.~(2), which for high $f_{d}$ goes as 
$ f_{p} \approx f_{d}$, in agreement with the phase diagram. 
Similarly, the  III--IV boundary follows Eq.~(3), which is 
also linear for large $ f_{d}$.  
The onset of region V also goes as $ f_{p} \propto f_{d}$. 

Next we vary the commensurability, from $ B/B_{\phi} = 0.75$ 
to $1.7$, producing the phase diagram in Fig.~3(b). Just 
above $B/B_{\phi} = 1$, the five phases of Fig.~1 are present. 
As $B/B_{\phi}$ is increased, the disordered-flow region III 
grows while the ordered-flow regions II and IV shrink. This 
is expected since an increase in $B/B_{\phi} $ effectively 
introduces more disorder via the addition of more interstitial 
vortices.   
For $ B/B_{\phi} > 1.3 $, the flow becomes more disordered, 
the II--III phase boundary becomes ill-defined, and the flow 
in region V is no longer composed of 1D incommensurate flow 
along the pinning rows but has a number of vortices flowing 
between the pinning rows.
At $ B/B_{\phi} = 1$, the commensurate case, we find only 
two phases: pinned and flowing, with the onset of flow 
occurring at $ f_{d} \approx f_{p}$. The vortex flow for the 
commensurate case is {\it elastic} since, unlike the case for 
$ B/B_{\phi} > 1$, there are no discommensurations that cause 
certain rows to move faster. For $B/B_{\phi} < 1$, where a 
number of vacancies appear in the vortex lattice, we observe 
a new {\it vacancy} flow phase, marked region VI.  The depining 
force for the onset of vacancy motion is considerably higher 
than that for  the onset of interstitial motion, in agreement 
with experiments \cite{Harada}.  Flow in V' is like in V, 
but now the faster-moving rows have vacancies. 

Figure 3(c) shows  a phase diagram in which the positions of 
the pinning sites are gradually disordered by randomly 
displacing them up to an amount $\delta r$ from the ordered 
pinning lattice position. In terms of the pinning lattice 
constant $ a$, we consider the  case $ \delta r = a/2$ to be 
a good approximation to a random pinning array.   The 
disordered region III grows and dominates the phase diagram 
for disorder greater then $ \delta r = a/6$, so that only three 
phases occur, in agreement with other simulations of random 
pinning arrays \cite{Koshelev,Kyu,Middleton}. 

Figure 3(d) shows the phase diagram for increasing both vortex 
$n_{v}$ and pin $n_{p}$ density, determined by $ B_{\phi}$. 
Here, the ordered-flow phase IV grows while the disordered flow 
region III shrinks. Increasing $ B_{\phi}$, increases $n_{v}$ 
and thus effectively decreases $f_{p}$. As seen in Fig.~3(a), 
this suppresses II and III, while favoring IV.  Phase diagrams 
for varying pinning radius $r_{p}$, angle of drive, temperature, 
and triangular pinning arrays will be presented and discussed 
in detail elsewhere \cite{CR}.

In conclusion, we have demonstrated several novel dynamical 
phases which are very distinct from those found in random 
pinning arrays. These phases are marked by pronounced changes 
in the $V(I)$ curves and noise fluctuations which should be 
very accessible experimentally. 
We have directly related these features to the pronounced 
changes in the vortex lattice structure and flow pattern. 
These phases result from the coupling between two species 
of vortices.  We have shown that certain phases show strong 
hysteresis while others do not. The dependence of these 
phases on various pinning and system parameters has been 
extensively studied and summarized in a series of phase 
diagrams \cite{CR}.   We hope that these results encourage 
the experimental search of these new dynamic phases.

Computing services were provided by the University of Michigan 
Center for Parallel Computing, partially funded by 
NSF grant CDA-92-14296.

\begin{figure}
\centerline{
\epsfxsize=3.5in
\epsfbox{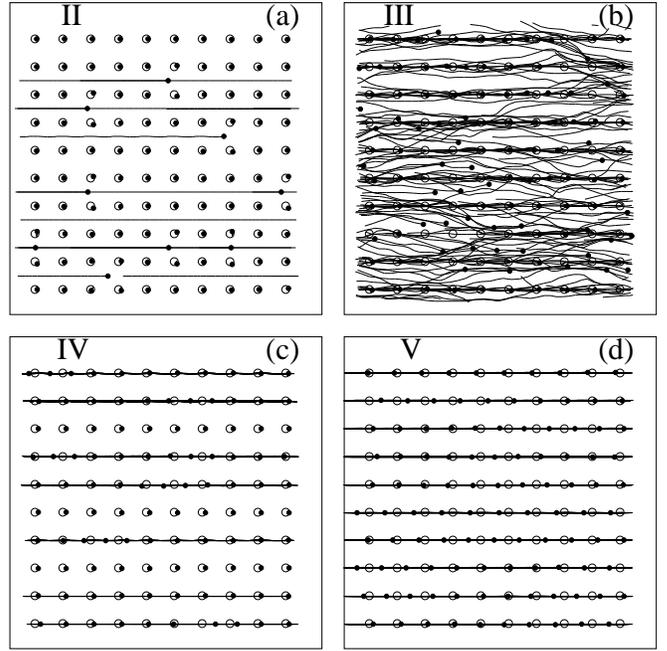}}
\caption{
Trajectories of the east-bound flowing vortices, for the 
voltage-current curve in Fig.~1, for regions:  II, 
interstitial flow (a); III, disordered flow (b); IV, 
incommensurate 1D flow (c); and V, moving (in)commensurate 
rows (d). 
The vortices are represented by black dots and pinning sites by open
circles. For clarity, only a $20\lambda \times 20\lambda $ subset 
of the $36\lambda \times 36\lambda$ sample is shown.
}
\label{fig2}
\end{figure} 

\begin{figure}
\centerline{
\epsfxsize=3.5in
\epsfbox{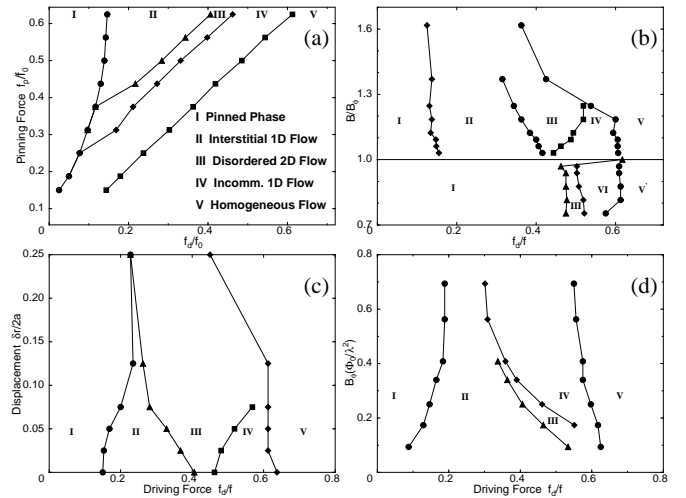}}
\caption{ 
Dynamic phase diagrams with a square pinning lattice.  Here, 
$B/B_{\phi} = 1.062$, $B_{\phi} = 0.25\Phi_{0}/\lambda^{2}$,
$r_{p} = 0.3\lambda$, and $f_{p} = 0.625f_{0}$, unless otherwise 
noted. 
(a) Pinning force $f_{p}$ versus driving force $f_{d}$. As 
$f_{d}$ is increased, the phase boundaries II--III, III--IV, 
and IV--V become linear. 
(b) $B/B_{\phi}$ versus $f_{d}$. For $B/B_{\phi} > 1$, regions 
I through V can be observed, with the disordered region III 
growing and the ordered-flow regions II and IV reducing in 
size. A similar situation occurs in (c) for gradually 
increasing the amount of disorder in the location of the pins. 
(d) Dynamic phase diagram for $B_{\phi}$ versus $f_{d}$. 
}
\label{fig3}
\end{figure}

\end{document}